\documentstyle[twocolumn,prb,aps,epsf]{revtex}
 
\begin{document} 
\twocolumn[
\title{Local modes, phonons, and mass transport in solid $^{4}$He. } 
\author{N. Gov and E. Polturak} 
\address{Physics Department,\\ 
Technion-Israel Institute of Technology,\\ 
Haifa 32000, Israel} 
\maketitle
\tightenlines
\widetext
\advance\leftskip by 57pt
\advance\rightskip by 57pt
 
\begin{abstract} 
We propose a model to treat the local motion of atoms in solid $^{4}$He as a 
local mode. In this model, the solid is assumed to be described by the Self 
Consistent Harmonic approximation, combined with an array of local modes. We 
show that in the bcc phase the atomic local motion is highly directional and 
correlated, while in the hcp phase there is no such correlation. The 
correlated motion in the bcc phase leads to a strong hybridization of the 
local modes with the T$_{1}(110)$ phonon branch, which becomes much softer 
than that obtained through a Self Consistent Harmonic calculation, in 
agreement with experiment. In addition we predict a high energy excitation 
branch which is important for self-diffusion. Both the hybridization and the 
presence of a high energy branch are a consequence of the correlation, and 
appear only in the bcc phase. We suggest that the local modes can play the 
role in mass transport usually attributed to point defects (vacancies). Our 
approach offers a more overall consistent picture than obtained using 
vacancies as the predominant point defect. In particular, we show that our 
approach resolves the long standing controversy regarding the contribution 
of point defects to the specific heat of solid $^{4}$He. 
\end{abstract} 
 
\vskip 0.2cm
]

\narrowtext
\tightenlines

\vspace{.2cm}

\section{Introduction} 
 
Atomic motion in solid ${\rm ^{4}He}$ was extensively studied over the 
years, both theoretically and experimentally\cite{glyde}. The quantum nature 
of the solid comes into play mainly through the large amplitude of zero 
point motion of the atoms in the lattice, an effect especially evident in 
the solid at its lowest density. In this paper, we shall discuss only these 
low density solids ($\sim $ 21 cm$^{3}$ molar volume). Calculations of the 
phonon spectrum within the self consistent harmonic approximation seem to 
agree fairly well with experiment in the hcp phase\cite{mink1}, while in the 
bcc phase, only the energy of the ${\rm T_{1}(110)}$ phonon is overestimated 
by a factor of two\cite{mink2}. Corrections resulting from the local motion 
of the atoms due to quantum effects are treated using correlated basis wave 
functions\cite{glyde}. A problem which so far was not addresed within these 
models is that of mass transport, arising from atomic exchange at low 
temperature. At higher temperatures, this exchange is assumed to occur with 
thermally activated point defects, traditionally taken as vacancies. With 
point defects, some additional complexity is to be expected, as here one 
deals with objects of a size similar to that of the zero point amplitude of 
the atom. The existence of thermally activated non-phonon excitations is 
well established experimentally by several different techniques, such as NMR%
\cite{allen,schuster,miyoshi,grigorev}, mass diffusion\cite{dyumin,berent}, 
ion mobility\cite{dahm}, and X-ray diffraction\cite{simmons}. However, the 
properties of point defects deduced from these various experiments have so 
far led to controversial conclusions. For example, taking the experimentally 
determined activation energies at face value, the calculated density of 
vacancies is so large that their contribution to the specific heat should be 
comparable to that of the phonons\cite{burns,dyumin}. This is not borne out 
by the specific heat data. Several attempts were made to reconcile this 
controversy by using the assumption that vacancies in solid ${\rm ^{4}He}$ 
are delocalized and occupy an energy band\cite{andreev,burns,dyumin}. In 
this approach, the magnitude of the specific heat is divided by the 
bandwidth, so that it can be made small through the choice of a 
suffficiently large bandwidth. However, in order to obtain a realistic 
magnitude of the specific heat one has to chose an unphysically large 
bandwidth, so that this approach is unsuccessful in settling the controversy%
\cite{dyumin}. 
 
In this paper, we propose a somewhat different description of the solid, one 
which treats the motion of the atom inside it's potential-well as a ''local 
mode''. Local modes as excitations in anharmonic crystals were discussed 
previously by Sievers and Takeno\cite{sievers}, and by Flynn and Stoneham%
\cite{flynn}. We show below that the other degrees of freedom are largely 
decoupled from this local motion. As such, this mode can be assumed to be an 
independent degree of freedom of the solid. The excited states of this local 
mode can be thermally activated, and seems to reproduce all the physical 
effects usually attributed to vacancies. Moreover, we show that this picture 
has none of the inconsistencies connected with vacancies. The properties of 
the local modes are different in the bcc and the hcp solid phases. In the 
bcc phase, we show that it is energetically advantegous for the local motion 
of the different atoms to be correlated. The new ground-state has several 
new features, among them a softened transverse phonon mode and a high-energy 
excitation branch which opens an additional channel for atomic diffusion. In 
contrast, the triangular symmetry of the hcp lattice frustrates the 
possibility of long-range ordering of the local modes, which therefore 
remain uncorrelated. In both phases the local modes are highly directional 
due to the anisotropic potential seen by an atom in the crystal, a 
constraint which we show leads to a small contribution to the specific heat 
of the solid. 
 
\section{Local modes in bcc $^{4}$He} 
 
Calculations of the ground-state energy of bcc $^{4}$He usually employ 
variational wavefunctions that account for the short-range correlations 
between the atoms \cite{glyde}. Since we want to focus here on the 
nature of the local motion of the atom inside the lattice, we will 
take a different approach. In this approach we would like to isolate the 
lowest energy excited state of the atom inside its potential well, and treat 
it as a local excitation of the lattice. This excited state consists of a 
local oscillatory motion of the atom along a particular direction and 
produces an oscillating electric dipole, similarly to the usual Van-der 
Waals interaction. Unlike in the case of the Van-der Waals interaction, in 
which the dipolar fluctuations are random, we show that in the bcc solid 
these local dipoles are correlated and a new ground-state of lower energy is 
created. We therefore begin by investigating the potential well of an atom 
in the bcc lattice. This can be done in the following way: Using the 
standard helium pair-potential $\upsilon (r)$ \cite{glyde} and taking the 
atoms as stationary we can map the potential-well near an atom along any 
direction in the lattice. In Fig.1 we plot this potential along the main 
directions (100),(110) and (111). It is clear that in the directions normal 
to the cube's faces (i.e. (100),(010) etc.) the confining potential well is 
very wide with a pronounced double-minimum structure. 
 
We also plot in Fig.2 the lowest two energy levels of a one-dimensional 
Schrodinger's equation for a $^{4}$He atom solved in each of the potential 
wells. The energy difference between these two levels is lowest in the (100) 
direction, of the order of $10{\rm K}$. The atomic displacement is described 
by a mixing between the lowest two energy levels in the potential well, 
which corresponds to a motion with amplitude of $\sim 1{\rm \AA }$ (in the 
(100) direction). Although the above treatment assumes that other atoms are 
stationary, it indicates the directions along which there will be large 
amplitude local motion in the solid. It is evident that the motion will 
be highly directional, with the largest amplitude along the (100) direction, 
while in the other directions the amplitude is much smaller. Allowing other 
atoms to move as well, so that they get out of each other's way, can only 
soften the potential well and reduce the value of the energy level 
difference, since the potential will effectively become shallower. For 
example, a calculation which allows opposite atoms to move slightly, reduces 
the energy difference in the (100) direction to almost $6{\rm K}$. 
 
In order to obtain meaningful numerical values, we scaled the potential $%
\upsilon (r)$ so that the calculated one-dimensional potential energy and 
kinetic energy, averaged over the different directions, reproduce the known 
kinetic, potential and total energy of the bcc phase \cite{glyde}: $%
\left\langle K.E.\right\rangle \simeq 34$K, $\left\langle V\right\rangle 
\simeq -40$K, $\left\langle E_{total}\right\rangle \simeq -6$K. 
 
Based on the above calculation, we shall assume that the atoms have a 
local-mode that is highly directional along one of the directions equivalent 
to (100), and of energy $6-10$K. Experimental evidence for the existence of 
such a ''local mode'' comes from NMR measurements of the linewidth (${\rm %
1/T_{2}}$) in bcc $^{3}$He-$^{4}$He mixture crystals\cite{schuster}. The 
motional narrowing of the NMR resonance line with temperature shows 
thermally activated behaviour with an activation energy of 7$\pm 1$K\cite 
{schuster,allen}. At temperatures above 1K, the NMR line in the solid 
becomes narrower than that of the liquid, indicating that atomic motion in 
the solid is faster than in the liquid. At the same time, the diffusion 
coefficient of the solid remains several orders of magnitude smaller than 
that of the liquid\cite{schuster}, indicating that this rapid motion is of a 
local nature. We propose to identify this rapid motion as associated with 
the excited state of an atom in the well, with an activation energy of $7\pm 
1$K. 
 
The highly directional motion of the atom leads to the creation of a local 
(oscillating) electric dipole in the direction of the motion. This local 
electric-dipole is created due to inertia, as the electronic cloud can be 
thought of as being slightly displaced relative to the ion. This electric 
dipole due to mixing of the $\left| s\right\rangle $ and $\left| 
p\right\rangle $ electronic-levels of the $^{4}$He atom. The amount of 
mixing can be estimated from perturbation theory as  
\begin{eqnarray} 
\psi &=&\left| s\right\rangle +\lambda \left| p\right\rangle \Rightarrow 
E_{0}\simeq \left\langle \psi \left| E\right| \psi \right\rangle 
-\left\langle s\left| E\right| s\right\rangle \simeq \lambda 
^{2}\left\langle p\left| E\right| p\right\rangle  \nonumber \\ 
\ &\Rightarrow &\lambda ^{2}\simeq 7/2.46\cdot 10^{4}\simeq 0.00284,\lambda 
\simeq 0.0168  \label{lamda} 
\end{eqnarray} 
 
where $\left| s\right\rangle $ and $\left| p\right\rangle $ stand for the 
ground-state and first excited-state of the $^{4}$He atom, $\lambda $ is the 
mixing coefficient and $\left\langle p\left| E\right| p\right\rangle \simeq 
2.46\cdot 10^{4}$ K is the excitation energy of the first atomic excited-state \cite 
{white}. The estimated mixing is small and the magnitude of the induced 
dipole moment is therefore  
\begin{equation} 
\left| {\bf \mu }\right| =e\left\langle \psi \left| x\right| \psi 
\right\rangle \simeq 2e\lambda \left\langle s\left| x\right| p\right\rangle 
\simeq e\cdot 0.03{\rm \AA }  \label{mu} 
\end{equation} 
 
where $\left\langle s\left| x\right| p\right\rangle \simeq 0.9{\rm \AA }$ . 
The estimation of the mixing $\lambda $ and the dipole-moment $\left| {\bf %
\mu }\right| $ serves to set an upper bound on the magnitude of this effect. 
Since the atoms possess an oscillatory electric-dipole moment they have 
long-range dipole-dipole interactions. The instantaneous dipolar interaction 
energy is given by  
\begin{equation} 
E_{dipole}=-\left| {\bf \mu }\right| ^{2}\sum_{i\neq 0}\left[ \frac{3\cos 
^{2}\left( {\bf \mu }\cdot \left( {\bf r}_{0}-{\bf r}_{i}\right) \right) -1}{%
\left| {\bf r}_{0}-{\bf r}_{i}\right| ^{3}}\right]  \label{edipole} 
\end{equation} 
 
where the sum is over all the atoms in the lattice, ${\bf r}_{i}$ being the 
instantaneous coordinate of the $i$-th atom. For oscillating dipoles with random phases,
the average instantaneous interaction energy summed over the lattice would be zero. 
However, 
the energy of a dipolar array can be made lower by correlating the phases of 
the oscillating atoms. The lowest interaction energy arrangement of the dipoles in the 
bcc lattice is such that they are oscillating with the same phase. Since the direction of the 
dipole shows the instantaneous direction of the motion or displacement, such 
a state is just a uniform motion or translation of the entire lattice. This 
arrangement is therefore unphysical, and we have to look for symmetric 
arrangements with respect to the number of up/down dipoles. The two 
arrangements shown in Fig.3 are the two 'antiferroelectric' configurations along the symmetry axes of the crystal
with individual dipoles oriented along the (001) direction, and a zero total 
dipole moment. For these two physical possibilities the sum in (\ref{edipole}%
) with a unit dipole is given in Fig.3. Thus, the ground state in our 
picture has the atoms executing this local oscillation in a correlated 
fashion, as shown in Fig.3a along one particular direction. Similar correlated motion exists in the other two orthogonal directions.
 
\section{Elementary excitations of the dipole ground-state} 
 
The ground state of the dipoles described in the preceding section will be 
affected by the excitations of the lattice, namely phonons. In fact, our 
basic assumption in which the local motion can be separated from these other 
degrees of freedom needs justification. The oscillatory atomic motion 
induced by the phonons will modulate the relative phases of the dipoles. Let 
us look at the ground state of the dipoles concentrating on oscillations
oriented along the (001) direction.
We now need to consider only phonons which will modulate the local motion along 
this direction. In the bcc structure, only 3 phonons fulfill this condition: 
L(001), T(100) and T$_{1}$(110). Let us calculate the energy of the dipolar 
array when modulated along these 3 directions. For a modulation along some 
direction ${\bf k}$ , the dipolar interaction energy is given by\cite{heller}%
:  
\begin{eqnarray} 
X\left( {\bf k}\right) &=&%
-\left| {\bf \mu }\right| ^{2}\sum_{i\neq 0}\left[%
\frac{3\cos ^{2}\left( {\bf \mu }\cdot \left( {\bf r}_{0}-{\bf r}_{i}\right)%
\right) -1}{\left| {\bf r}_{0}-{\bf r}_{i}\right| ^{3}}\right] \nonumber \\
&&\ \ \exp \left[ 
2\pi i{\bf k}\cdot \left( {\bf r}_{0}-{\bf r}_{i}\right) \right]  \label{xk}
\end{eqnarray} 
 
At $k=0$ the interaction matrix $X(k)$ is just the dipolar energy (\ref 
{edipole}). 
 
In Fig.4 we plot the value of $X(k)$, the energy of the dipolar array 
modulated by the relevant phonons:\ L(001), T(100), and T$_{1}$(110), for 
dipole moment $\left| {\bf \mu }\right| =1$. We see that for a modulation by 
L(001) and T(100) the periodicity of $X(k)$ is over a full unit-cell, that 
is twice the periodicity of these phonons. Since symmetric functions of 
periodicities $\pi /a$ and $2\pi /a$ are orthogonal, the wavefunctions of 
the phonons and of the dipole-excitations are orthogonal along these 
directions. The dipole array cannot therefore be excited by any phonon along 
these two directions. Regarding the modulation by the T$_{1}$(110) mode, 
here the periodicity of $X(k)$ is the same as that of the T$_{1}$(110) 
phonon, which can therefore couple to the dipole array. We conlude therefore 
that the coupling of the local modes to the lattice excitations is limited 
to a single phonon mode, justifying our assumption that the local modes can 
be treated separately to a good approximation. Thus, the only elementary 
excitations of the dipole array would be in the (110) direction, in the form 
of the T$_{1}$(110) phonon. We shall now calculate the dispersion relation 
of such an excitation by a mean-field solution of an effective Hamiltonian. 
 
The Hamiltonian treatment of interacting local excitations was developed 
originally by Hopfield \cite{hopfield} for the problem of excitons in a 
dielectric material. The local excitations are treated as bosons and the 
effective Hamiltonian describing their behavior is \cite{anderson} 
 
\begin{eqnarray} 
{H_{loc}} &=&{\sum_{k}}(E_{0}+X(k))\left( {{b_{k}}^{\dagger }}{b_{k}}+{\frac{1}{%
2}}\right) \nonumber \\
&&\ \ +{\sum_{k}}X(k)\left( {{b_{k}}^{\dagger }}{b_{-k}^{\dagger }}%
+b_{k}b_{-k}\right)  \label{hloc} 
\end{eqnarray} 
where ${{b_{k}}^{\dagger },}{b_{k}}$ are creation/anihilation operators of 
the local mode, and $X(k)$ is the interaction matrix element given above (%
\ref{xk}). The Hamiltonian ${H_{loc}}$ (\ref{hloc}) which describes the 
effective interaction between localized modes can be diagonalized using the 
Bogoliubov transformation ${\beta _{k}}=u(k)b_{k}+v(k)b{^{\dagger }}_{-k}$. 
The two functions $u(k)$ and $v(k)$ are given by:  
\begin{equation} 
{u^{2}}(k)={\frac{1}{2}}\left( \frac{E_{0}{+X(k)}}{{E(k)}}+1\right) ,{v^{2}}%
(k)={\frac{1}{2}}\left( \frac{E_{0}{+X(k)}}{{E(k)}}-1\right)  \label{uv} 
\end{equation} 
 
Where $E(k)$, the energy spectrum of the diagonalized Hamiltonian is:  
\begin{equation} 
E(k)=\sqrt{E_{0}\left( E_{0}+2X(k)\right) }  \label{ek} 
\end{equation} 
 
The ground-state wavefunction of the local-modes is given by \cite{huang}:  
\begin{equation} 
\left| \Psi _{0}\right\rangle =\prod_{k}\exp \left( \frac{v_{k}}{u_{k}}{{%
b_{k}}^{\dagger }}{b_{-k}^{\dagger }}\right) \left| vac\right\rangle 
\label{psi0} 
\end{equation} 
 
We show in Fig.4 $X(k)$ in the (110) direction for a unit dipole moment. We 
now need to fix the size of the dipole moment $\left| {\bf \mu }\right| $ in 
order to calculate the energy spectrum. We would like, according to our 
definition ot the local mode, that the energy cost of flipping the direction 
on a single dipole out of the ground state to be $E_{0}$. This condition is 
equivalent to demanding that $2\left| X(k=0)\right| =E_{0}$. We also see 
from (\ref{ek}) that in order for the dipoles to have a gapless mode at $%
k\rightarrow 0$ we must have that: $X(k=0)=-E_{0}/2$. Using this condition, 
the value of $E_{0}=7$K (see previous section) determines the size of the dipole-moment as: $\left|  
{\bf \mu }\right| \simeq e\cdot 0.01{\rm \AA }$. This value is indeed 
smaller than our previous estimation, which was an upper bound on the size 
of dipole moment (\ref{mu}). 
 
From its very definition, the phase modulation in the (110) direction of the 
atomic motion with energy $E(k)$ (Eq.\ref{ek}) should be just the T$_{1}$%
(110) phonon. In Fig.5 we compare the experimental values of T$_{1}$(110) 
taken from neutron scattering with the calculated $E(k)$. The agreement is 
excellent for all $k$. From (\ref{ek}) and Fig.4 we find that at the edge of 
the Brillouin zone the energy $E(k)$ of the phonon is just the bare energy 
of the local mode, $E_{0}$, since $X(\sqrt{2}\pi /a)=0$ and the dipoles have 
changed from the configuration of Fig.3a to Fig.3b. We recall that the value 
of $E_{0}$ that we used was taken from NMR data. The agreement between these 
two independent determinations of $E_{0}$, that from NMR and that from 
neutron scattering, emphasizes the self-consistency of our description. We 
stress that the value of $E_{0}$ is the only empirical input into the 
calculation, while the functional behavior is completely defined by the 
lattice structure and the dipolar form of the interactions. 
 
Our model indicates that only the T$_{1}$(110) phonon would be different 
than obtained from the SCH calculation, because it is the only excitation 
which couples to the local motion. Indeed, in the experiment\cite{mink2} 
this is the only phonon branch which is not described well by the SCH 
calculation. The fact that the SCH calculation works rather well for other 
directions, is consistent with our picture in which there are no elementary 
excitations of the dipole array in these directions. 
 
\section{Harmonic and local-mode hybridization} 
 
An equivalent way to describe the mutual influence between the local modes 
and the lattice can be done by taking the interaction between the dipoles as 
resulting from an exchange of virtual harmonic transverse fluctuations of 
the lattice. An analogous case, that of excitons in a dielectric, was 
treated by Hopfield\cite{hopfield} (in that case, the interaction is 
mediated through the exchange of virtual photons). The two excitations, the 
local mode and virtual harmonic phonon, are then hybridized through the same 
dipolar interaction matrix $X(k)$, which we used in the direct interaction 
picture. In our case, the natural choice for the mediating virtual phonon is 
the T$_{1}$(110) phonon, as calculated by the SCH method. The use of the 
phonon calculated by the SCH method is important, since this calculation is 
largely independent of the local degrees of freedom described by the local 
mode. The motivation for using this approach is that it allows us to obtain 
an additional branch of the excitation spectrum which has observable 
consequences. 
 
The Hamiltonian describing the local-mode and SCH-phonon hybridization is  
\cite{hopfield}  
\begin{equation} 
H=H_0+H_{loc}^0+H_c  \label{hhyb} 
\end{equation} 
 
where the three components of (\ref{hhyb}) are: 
 
The SCH-phonon 
 
\begin{equation} 
H_0=\sum_k\varepsilon (k)a_k^{\dagger }a_k  \label{h0} 
\end{equation} 
 
where $a$ are Bose operators and $\varepsilon (k)$ is the SCH-phonon 
spectrum. 
 
The local-mode  
\begin{equation} 
{H_{loc}^{0}}={\sum_{k}}E_{0}{{b_{k}}^{\dagger }}{b_{k}}  \label{hloc0} 
\end{equation} 
 
The part describing the first order (dipolar) coupling between the phonons 
and the localized modes is ~\cite{hopfield}  
\begin{eqnarray} 
H_{c} &=&{\sum_{k}}\left( \lambda (k){b_{k}}+\mu (k){a_{k}}\right) ({{a_{k}}%
^{\dagger }}+{a_{-k}})  \label{hc} \\ 
&&\ \ -\left( \lambda (k){{b_{k}}^{\dagger }}-\mu (k){{a_{k}}^{\dagger }}%
\right) ({a_{k}}+{{a^{\dagger }}_{-k}})  \nonumber 
\end{eqnarray} 
 
where, in the dipolar approximation (and a cubic lattice) the two functions $%
\lambda $ and $\mu $ are given by $\lambda (k)=iE_{0}\left( -{\frac{3X(k)}{{%
2\epsilon (k)}}}\right) {^{\frac{1}{2}}}$ and $\mu (k)=-E_{0}\frac{{3X(k)}}{{%
2\epsilon (k)}}$, with ${X(k)}$ the dipole matrix element (\ref{xk}). This 
is just the standard coupling Hamiltonian of an atom to a transverse photon 
field, which is replaced here by the T$_{1}$(110) SCH phonon. The total 
Hamiltonian $H$ does not involve quartic terms and can be diagonalized using 
the canonical transformation \cite{hopfield}  
\begin{eqnarray} 
\alpha _{1} &=&Aa_{k}+Bb_{k}+C{a^{\dagger }}_{-k}+D{b^{\dagger }}_{-k}  
\nonumber  \\ 
{\alpha }_{2} &=&Ba_{k}+Ab_{k}+D{a^{\dagger }}_{-k}+C{b^{\dagger }}_{-k} 
\label{alfa} 
\end{eqnarray} 
where these operators describe the two branches of the hybridized energy 
spectrum. The transformation functions $A(k),B(k),C(k),D(k)$ can be written 
down explicitly\cite{hopfield}. The corresponding dispersion relation is  
\begin{equation} 
\frac{{\epsilon ^{2}}(k)}{{{E^{2}}(k)}}=1-\frac{6}{E_{0}}\frac{X(k)}{{1-}%
\left( \frac{E(k)}{E_{0}}\right) ^{2}}  \label{ehop} 
\end{equation} 
 
which describes two energy branches $E_{1}$ and $E_{2}$. The equivalence of 
this and the treatment in the previous section is due to the use of the same 
dipole interaction matrix $X(k)$ in both. With $X(k)$, we can solve (\ref 
{ehop}) to find the two energy branches:  
\begin{eqnarray} 
E_{1} &=&\varepsilon (k)/2  \nonumber \\ 
E_{2} &=&2E_{0}  \label{e2} 
\end{eqnarray} 
 
We see that the energy of the lower branch, $E_{1}$ is half of that of the 
SCH-phonon for all $k$. Comparing the SCH calculation and the experimental 
results for the T$_{1}$(110) phonon, we find that there is indeed a 
constant, $k$ independent ratio between them. This ratio turns out to be $%
\sim 1.7$ for all momenta, close to the predicted ratio of 2 (\ref{e2}) (Fig.6). 
This small discrepancy is within the uncertainty of the SCH calculations and 
the experimental data. Despite this small discrepancy we find the agreement 
very satisfying. Our picture of a hybridization of a dipole array with the 
harmonic lattice can therefore account for the main effect of the local 
motion. The upper branch, $E_{2}$, a dispersionless excitation with energy 
of 14K (Fig.6), was possibly observed in inelastic neutron scattering experiment\cite 
{osgood}, in which a strong feature was observed at an energy transfer of 
1.4meV. This feature was interpreted by Glyde\cite{glyde} in terms of an 
interference effect between phonon modes, which is not inconsistent with our 
hybridization picture. Another possible way to observe this excitation would 
be by Raman scattering. In addition, this energy branch is important in the 
process of mass diffusion, and will be discussed in this context in a 
following section. 
 
We would like to point out that local modes can arise also in classical 
crystals, as a result of a large anharmonicity\cite{sievers}. As shown by 
these authors, these local modes can assume many of the roles of vacancies. 
There are several similarities between this work and ours; first, the 
treatment in non-perturbative; second, in a simple cubic lattice, they also 
find that there are two excitation branches. These similarities occur despite the 
different basic assumptions of the classical model and the present work, 
which uses the quantum properties of the crystal as the starting point. 
 
\section{Local modes in hcp $^{4}$He} 
 
We now turn to discuss the effects arising from the local motion in the hcp 
structure. Repeating the calculation done in section II, we plot in Fig.7 
the potential well along the main directions of the hcp crystal, namely 
(1000),(0100) and (0001). We also plot in Fig.7 the lowest two energy levels 
of a one-dimensional Schrodinger's equation solved in each of the potential 
wells. The energy difference between these two levels is lowest in the 
(0100) direction, of the order of $16$K. Similarly to what was stated for 
the bcc structure, the calculation indicates the direction along which there 
will be large amplitude local motion. 
 
As in the bcc phase, evidence for the existence of a ''local mode'' comes 
from NMR measurements\cite{miyoshi,schuster2} of the linewidth (${\rm 1/T_{2}%
}$) with energy $\simeq 14$K. We propose therefore that the NMR experiments 
measure the energy of the bare local mode. 
 
In contrast to the bcc phase, we do not expect the local electric-dipoles in 
the hcp phase to have long-range order. This is due to the fact that there 
is geometric frustration against an 'antiferroelectric' order in a 
triangular lattice. We calculated the dipolar interaction energy for several 
simple dipole arrangements preserving the net zero dipole moment, and did 
not find any arrangement in which this energy was negative. Comparison of the 
SCH calculation of the phonons spectrum with experiment reveals that there 
is an overall good agreement, with no exceptions, such as found for the T$%
_{1}$(110) phonon in the bcc phase. Thus, we conclude that there is no long 
range order within the dipole array like in the bcc phase, and the local 
modes in the hcp solid are largely uncorrelated. 
 
\section{Atomic self-diffusion} 
 
We have seen that in the first excited state of the atom in its potential 
well, the atom oscillates with a comparatively large amplitude ($\sim 1{\rm %
\AA }$). This motion can have an effect on the atomic exchange rate which is 
governed by the overlap of the wave functions of neighbouring atoms. At low 
temperature, the atomic exchange, or self diffusion, occurs by tunneling and 
is small ( D$_{0}\sim $10$^{-9}$cm$^{2}/\sec $ in the hcp phase\cite{dyumin}%
). Thermally activated diffusion is traditionally attributed to vacancies. 
We would like to show that thermally excited local modes can produce self 
diffusion at rates which compare favorably with experimental data. 
 
The theoretical calculation of the zero-temperature exchange rate of atoms 
in solid Helium is a long-standing problem \cite{roger}. It is extremely 
difficult to perform accurate theoretical calculations of this effect due to 
the smallness of the exchange frequency as compared with the Debye 
frequency. In these calculations, the atoms participating in the exchange 
are treated as tunneling along a one-dimensional closed path. At high 
densities, it is found that the effective potential barrier for tunneling is 
mainly due to the potential energy associated with the elastic deformation 
of the crystal during the atomic exchange. At lower densities it is more 
difficult to calculate the height of the tunneling barrier, but empirically 
it turns out to be of the order of the kinetic energy of the atoms (30K to 
40K). Thus, in order to obtain a rough estimate of the exchange rate, it is 
sufficient to view the problem as that of a one-dimensional tunneling of 
free particles through a square barrier. Since we're dealing with low 
density solid, the height of the barrier should be comparable with the 
kinetic energy of the atoms. We would like to compare the exchange rate of 
atoms in the ground state with that of atoms in the excited state of the 
local mode. 
 
We begin with the hcp phase, looking at the (0100) direction, that of the 
largest amplitude atomic motion. We take the atom in the ground-state with 
it's kinetic zero-point energy ($E_{kin}\simeq 10$K) and adjust the barrier 
height to reproduce the experimental rate of exchange at low temperatures. 
We take the width of the barrier to be $a=$3.6${\rm \AA }$, i.e. the nearest 
neighbour distance. We find that a barrier height of 44K gives a 
transmission probability $\Gamma \simeq 10^{-7}$, and a self diffusion 
coefficient D$_{0}^{G}\simeq (E_{kin}/\hbar )a^{2}\Gamma \sim $10$^{-9}$cm$%
^{2}/\sec $ , which is consistent with the experimental results. The barrier 
height found in this estimation is consistent with the empirical value, of 
the order of the total kinetic energy of the atoms in the solid. We now turn 
to thermally activated self diffusion. In the first excited state of atoms 
in the well, the total energy of the atom is 24K (10K from the ground state 
+ 14K for the first excited level of the local mode). The barrier height 
remains unchanged, but the atom can now tunnel from one of the lobes of the 
excited-state wavefunction (see Fig. 2), which are approximately 1${\rm \AA } 
$ from its equilibrium position. Since it tunnels into an identical excited 
state, the effective width of the barrier is now $a=$3.6-2=1.6${\rm \AA }$. We thus 
find a transmission probability in the excited state $\Gamma \simeq 10^{-2}$%
, and a prefactor of the diffusion coefficient D$_{0}^{E}\simeq 
(E_{kin}/\hbar )a^{2}\Gamma \sim $10$^{-4}$cm$^{2}/\sec $. This means in our 
model that as atoms are thermally excited out of the ground state, the 
diffusion coefficient would increase exponentially with temperature as D$%
_{0}^{E}exp(-E_{0}/kT).$ The experimental values\cite{dyumin} are D$%
_{0}^{E}\approx $ 10$^{-4}$cm$^{2}$/sec, and that of the activation energy $%
E_{0}$=$13.9$K, which is very close to the value of $E_{0}=$14K we took from 
NMR as describing the energy of the local mode. We therefore conclude that 
exchange of atoms occupying a thermally activated local mode can account for 
self diffusion at a rate usually attributed to vacancies in the hcp phase. 
 
The rate of self diffusion in the bcc phase is an order of magnitude larger 
than in the hcp phase \cite{dyumin}. In this phase, the direction of largest 
local atomic motion is (100). Repeating the above calculation of the 
tunneling rate for this case, we have the ground-state kinetic zero-point 
energy similar to the hcp ($E_{kin}\simeq 10$K), and barrier width along the 
(100) direction of $a=$4.12${\rm \AA }$. For a tunneling rate at low 
temperatures which would be higher by an order of magnitude compared with 
the hcp phase we need a barrier height of $\sim 30$K, which is again close 
to the total kinetic energy of the atom in the ground-state ($\sim 34$K)\cite{glyde}%
. In order to look at thermally activated self diffusion, the model used 
above for the hcp phase is inapplicable, since the energy of the excited 
state $E_{2}=2E_{0}\simeq 14^{o}{\rm K}$ (\ref{e2}) is associated with more 
than one atom. In order to understand the physical nature of this excitation 
we plot $n_{b}(r)$ (Fig.8), the spatial extent of the density of atoms excited by 
this branch of the local-modes in the (110) direction. This is the Fourier 
transform of the $k$-space density of the localized-modes in the upper 
branch (\ref{alfa}):  
\begin{eqnarray} 
n_{b}(k)=\left\langle b_{k}^{\dagger }b_{k}\right\rangle _{2}=\left| 
C(k)\right| ^{2}  
\label{nb}
\end{eqnarray} 
 
What is seen in Fig.8 is that the function $n_{b}(r)$ extends over two unit 
cells in real space, in the (110) direction. Since the energy is twice the energy of the bare 
local-mode, it can be interpreted qualitatively as two atoms excited in 
adjacent unit cells, each having an energy $E_{0}=$7K. This corresponds to a 
''flip'' of the dipole moment of these two atoms. In Fig. 9 we show that with 
this excitation, there are 4 atoms oscillating in such a way, as to reduce 
the potential barrier for the exchange of atoms 1 and 2. This type of atomic 
exchange resembles phonon assisted tunneling, where the correlated 
''breathing'' motion of the 4 atoms is locally equivalent to a phonon 
excitation at the edge of the Brillouin zone. This type of self diffusion 
was considered in the past\cite{flynn}, and was recently found to be consistent with experiment
in bcc $^{4}$He\cite{berent} as the dominant mass transport 
channel, with an activation energy of 14.8K\cite{dyumin,berent}. Thus, 
altough quantitative calculations of the rate of difusion are outside the 
scope of this paper, both the mechanism and the activation energy are in 
accord with experiment. 
 
\section{Contribution of local modes to the specific heat} 
 
One of the controversial issues with vacancies in solid $^{4}$He is that 
based on the measured values of the activation energy, their estimated 
contribution to the specific heat is comparable to that of the phonons\cite 
{burns,dyumin}. Yet, there in no experimental evidence for it. Let us 
consider the contribution of the local modes presented here to the specific 
heat. 
 
Since the energy of the local mode is different in each direction of the 
lattice, the highest possible contribution to the specific heat would come 
from the excitation of local modes along the directions where the excitation 
energy is lowest. Because of the restricted solid angle in which these modes 
are active, the fraction of phase space occupied by them would be 
correspondingly small, and their contribution to the specific heat would be 
much reduced compared with that obtained previously. In these previous 
estimations, the point defect was assumed to occupy phase space uniformly, 
thus yielding a large contribution. To obtain a quantitative estimate of the 
above effect we calculate the ratio between the contribution to the specific 
heat from directional local modes and the total experimental specific heat 
of the bcc\cite{hoffer} and hcp\cite{gardner} solid phases. We used the 
energies calculated above for the excitation energies along the principal 
directions of the bcc and hcp lattice, and a simple linear interpolation 
for the excitation energies in the intermediate orientations. This ratio is 
plotted against temperature in Fig.10. For the hcp phase, the maximum 
contribution is less than 1\%, while the bcc phase it is $\sim 5$\%. The 
decrease of this ratio near T=1.65K is due to the premelting enhancement of 
the total specific heat\cite{hoffer}. These values fall within the error 
bars of the specific heat data, and explain why no thermally activated 
contribution is seen in solid $^{4}$He. 
 
\section{Conclusion} 
 
In this work we have proposed a new approach to treat the local behaviour of  
$^{4}$He atoms in the bcc and hcp solid phases. We treat the excitations of 
atoms inside their potential well as local modes. The anisotropy of the 
potential renders these modes highly directional. 
Due to the symmetry of the bcc phase we propose that the local-mode is 
hybridized with the harmonic density fluctuations (SCH). The hybridization 
is described by the dipole-dipole interaction and the spectrum of the 
hybridized T$_{1}(110)$ phonon is calculated. An additional excitation 
branch is identified and it is this branch which seems to control the 
anomalously large self-diffusion in the bcc solid. In the hcp phase, the 
symmetry does not allow for correlations of these local modes. Consequently, 
there is no hybridization with the phonons, and the thermally activated self 
diffusion in this phase is controlled by the energy of the bare local mode. 
 
The directionality of the local modes means that their contribution to the 
specific-heat of the solid is negligible. We therefore demonstrated that the 
local mode approach can describe experimental data coming from 
neutron-scattering, NMR and diffusion experiments within one physical model, 
while at the same time resolving a long-standing discrepancy concerning the 
specific heat contribution of point defects. The physical picture of our 
proposed local mode is very different from the classical picture of a 
vacancy. We therefore see no reason to consider point defects in solid $^{4}$%
He as vacancies, as they can be consistently treated as natural excitations 
of the solid lattice, without having to physically remove atoms from it. 
 
{\bf Acknowledgements} 
 
This work was supported by the Israel Science Foundation and by the Technion 
VPR fund for the Promotion of Research.

\newpage
\begin{figure}[tbp]
\input epsf \centerline{\ \epsfysize 9.5cm \epsfbox{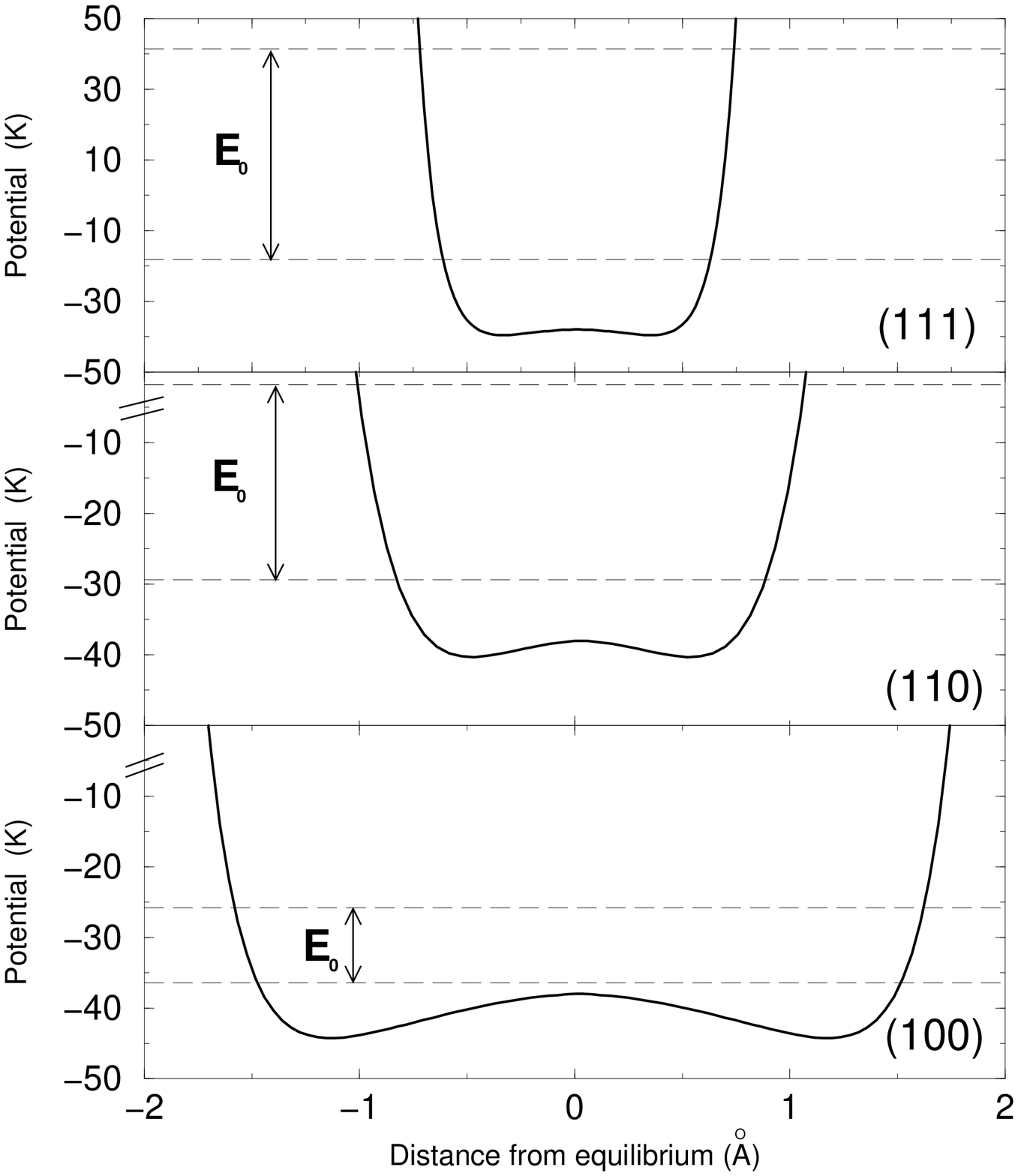}}
\caption{The potential-well of an atom in bcc $^{4}$He along different
directions. The energy difference $E_{0}$ between the lowest two energy
levels (dashed lines) are: (111)- $59.5$K, (110)- $27.6$K, (100)- $%
10.6$K.}
\end{figure}
\begin{figure}[tbp]
\input epsf \centerline{\ \epsfysize 9.5cm \epsfbox{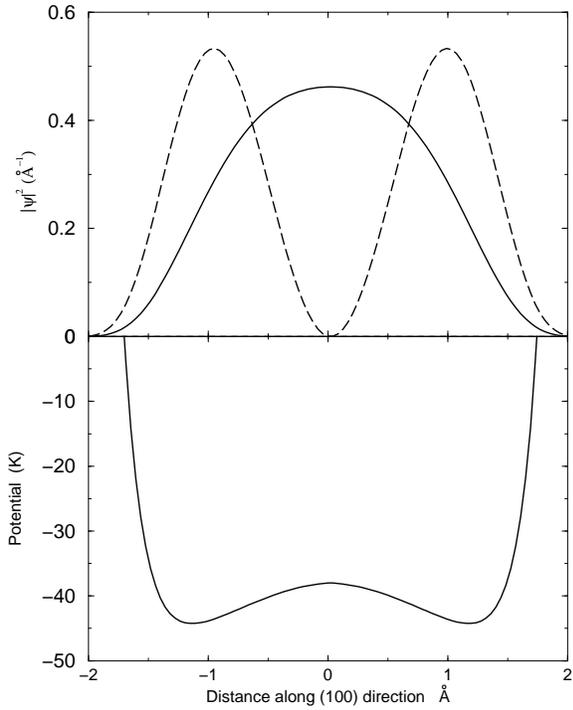}}
\caption{The probability distribution in the (100) potential well (shown
in the lower part). The groundstate wavefunction is the solid line while the
dashed line is the first excited state.}
\end{figure}
\begin{figure}[tbp]
\input epsf \centerline{\ \epsfysize 9.5cm \epsfbox{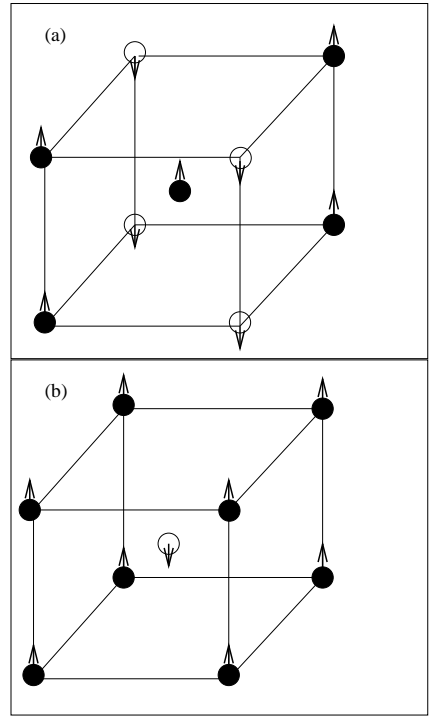}}
\vskip 8mm
\caption{The two 'anti-ferroelectric' arrangements of the dipoles lying along the
major axes of the bcc phase. The atoms having the same dipole moments have the
same shade. The sum of the dipole-dipole interaction (Eq.3) for a unit dipole-mo
ment
are: (a) -0.08, (b) 0.0 (${\rm \AA }^{-3}$).}
\end{figure}
\begin{figure}[tbp]
\input epsf \centerline{\ \epsfysize 11.0cm \epsfbox{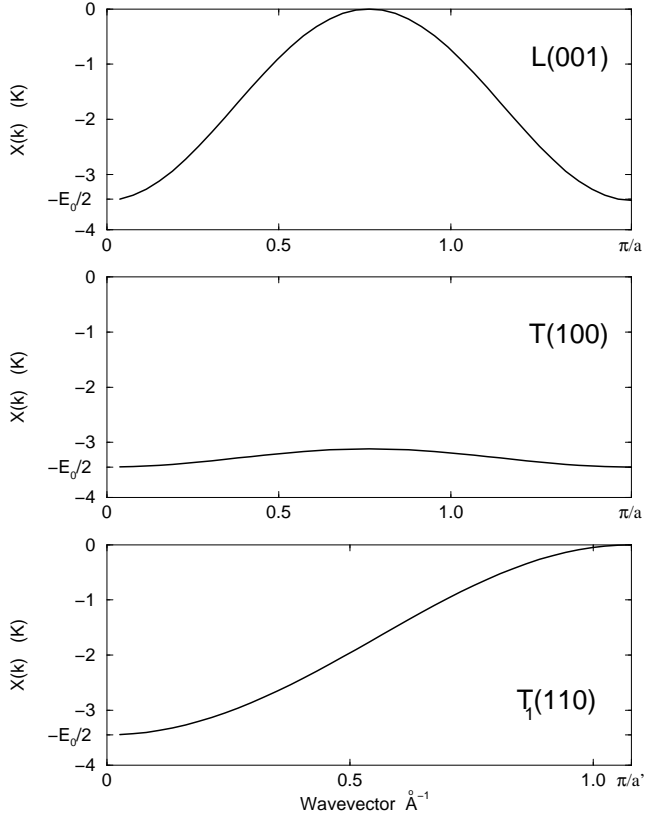}}
\caption{The calculated interaction matrix $X(k)$ (Eq.4) as a function of the
wavevector $k$, for the three phonon modes that could affect a dipolar
array. The dipole moment has been normalized to give a gapless mode: 
$X(k=0)=-E_{0}/2$. 
The unit-cell dimensions $({\rm \AA })$: $a=4.12/2$, $a'=a \protect{\sqrt{2}}$.}
\end{figure}
\begin{figure}[tbp]
\input epsf \centerline{\ \epsfysize 9.5cm \epsfbox{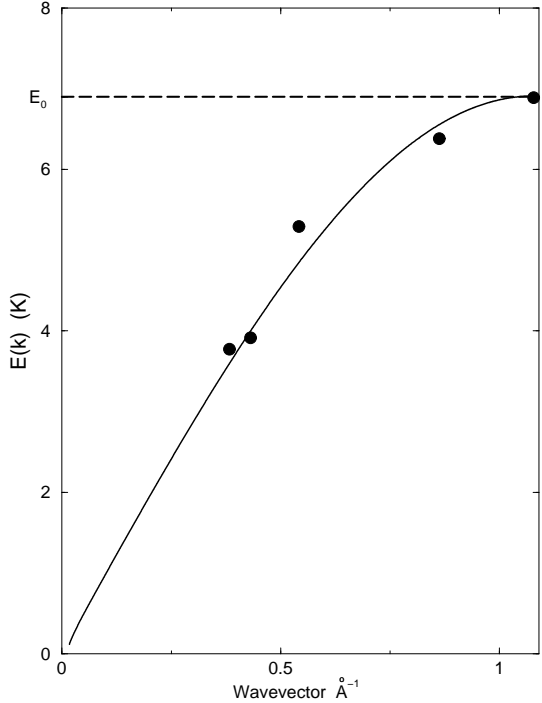}}
\caption{The experimental data [3] (solid circles) for the T$_{1}$(110)
phonon compared with the calculation (Eq.7) (solid line). Also shown is the
energy of the bare local-mode $E_{0}=7$K.}
\end{figure}
\begin{figure}[tbp]
\input epsf \centerline{\ \epsfysize 9.5cm \epsfbox{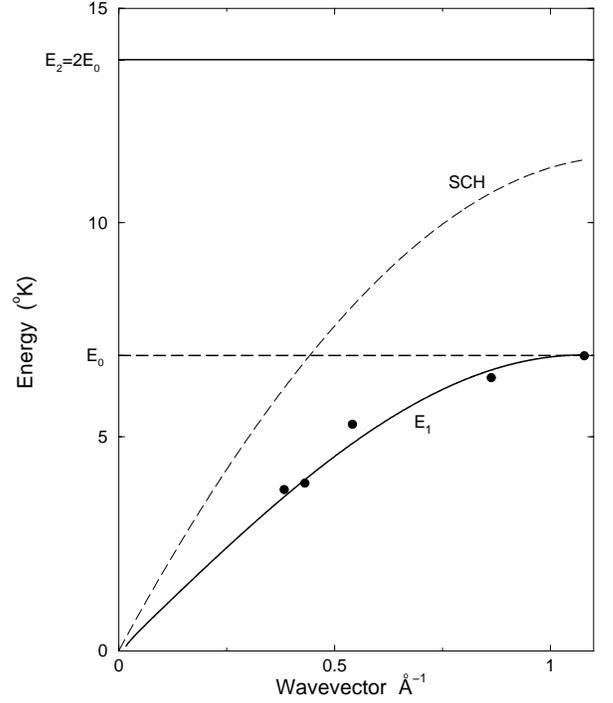}}
\caption{The SCH calculation [1] (dashed line) compared with the
experimental results [3] (solid circles) for the T$_{1}$(110) phonon. Solid line
 is our calculated spectrum (Eq.7). Also
shown is the prediction of the hybridization model (Eq.15) for the high-energy
branch $E_{2}=2E_{0}$.}
\end{figure}
\begin{figure}[tbp]
\input epsf \centerline{\ \epsfysize 9.5cm \epsfbox{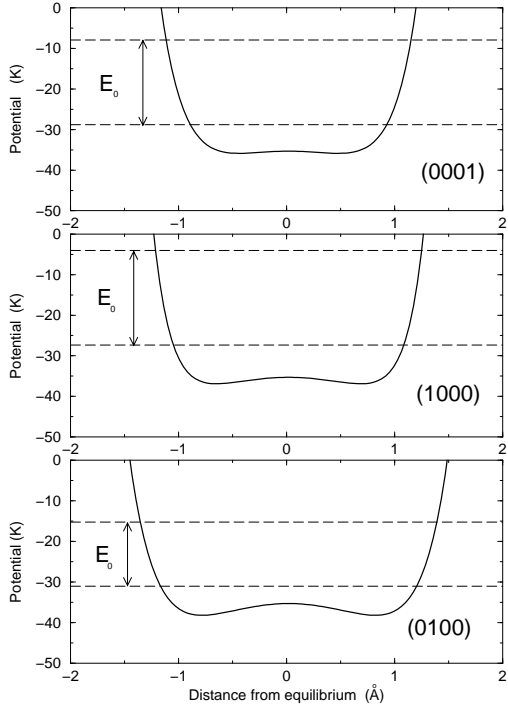}}
\caption{The potential-well of an atom in hcp $^{4}$He along different
directions. The energy difference $E_{0}$ between the lowest two energy
levels (dashed lines) are: (0100)- $16.0$K, (1000)- $21.0$K, (0001)- $%
23.0$K.}
\end{figure}
\begin{figure}[tbp]
\input epsf \centerline{\ \epsfysize 9.5cm \epsfbox{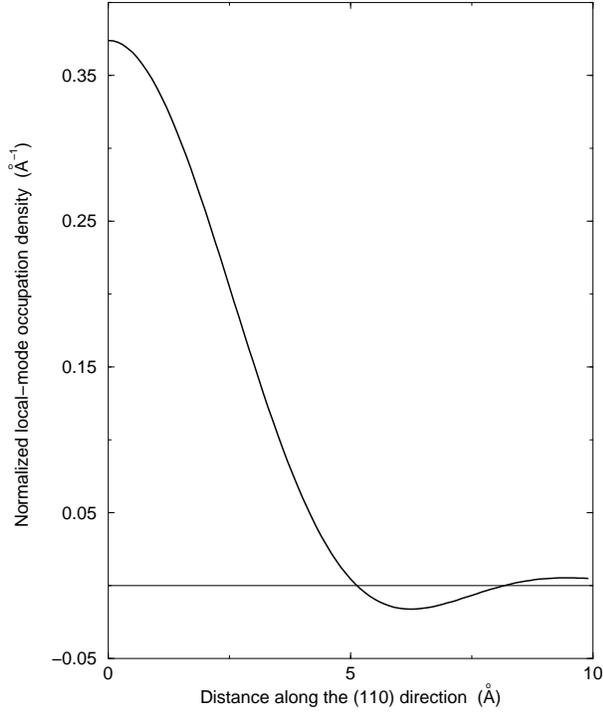}}
\caption{The normalized Fourier transform of the $k$-space density of the
local-mode occupation in the high-energy branch $E_{2}$ (Eq.16), along the (110)
 direction.}
\end{figure}
\begin{figure}[tbp]
\input epsf \centerline{\ \epsfysize 6.0cm \epsfbox{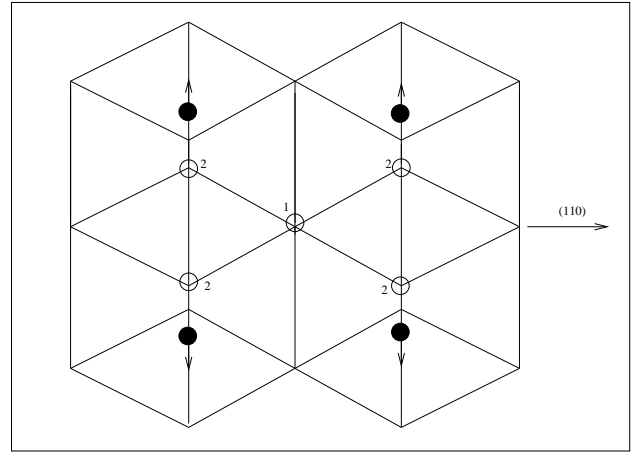}}
\vskip 8mm
\caption{Schematic picture of an excitation of the high energy branch $2E_{0}$%
, as two adjacent local modes along the (110) direction. The filled circles
represent atoms that execute the breathing motion, allowing atoms 1 and 2 (open
circles) to exchange places more easily.}
\end{figure}
\begin{figure}[tbp]
\input epsf \centerline{\ \epsfysize 9.5cm \epsfbox{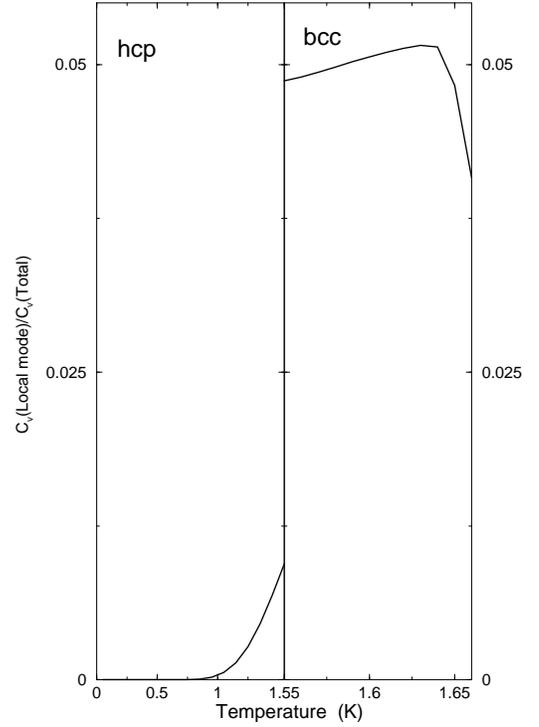}}
\vskip 8mm
\caption{The ratio of the contribution of the local
mode to the total specific heat in the hcp and bcc phases [24,25].
For clarity, the temperature scale for the bcc phase was expanded.}
\end{figure}

\end{document}